\documentclass[12pt]{article}

\usepackage{epsfig}
\usepackage{amsfonts,amssymb,newlfont}

\DeclareMathAlphabet{\mathitbf}{T1}{cmr}{bx}{it}

\begin{document}

\title{Low $T$ Dynamical Properties of Spin Glasses Smoothly
Extrapolate to $T=0$}

\author{Enzo Marinari$^{(a)}$, Giorgio Parisi$^{(a)}$ 
and Juan J. Ruiz-Lorenzo$^{(b)}$\\[0.3em]
$^{(a)}$  
{\small Dipartimento di Fisica, SMC and UdR1 of INFM  and INFN}\\ 
{\small Universit\`a di Roma {\em La Sapienza} }\\
{\small P. A. Moro 2, 00185 Roma, Italy}\\
{\small \tt enzo.marinari, giorgio.parisi@roma1.infn.it}\\[0.3em]
$^{(b)}$  
{\small Departamento de F\'{\i}sica, Facultad de Ciencias}\\
{\small Universidad de Extremadura}\\
{\small E-06071 Badajoz, Spain}\\
{\small \tt ruiz@unex.es}\\[0.5em]
}

\date{March 14, 2002}

\maketitle

\begin{abstract}
We compare ground state properties of $3D$ Ising Spin Glasses with
Gaussian couplings with results from off-equilibrium numerical
simulations at non zero (but low) temperatures.  We find that the
non-zero temperature properties of the system smoothly connect to the
$T=0$ behavior, confirming the point of view that results established
at $T=0$ typically also give relevant information about the $T\ne 0$
physics of the system.
\end{abstract}

\thispagestyle{empty}

\newpage

%%%%%%%%%%%%%%%%%%%%%%%%%%%%%%%%%%%%%%%%%%%%%%%%%%%%%%%%%%%%%%%%%%%%%%%%%%
\section{Introduction\label{S-INTRODUCTION}}

Spin glass physics is difficult. Because of how complex it is to
establish results that go further than the mean field level (see for
example \cite{BOOKS} and references therein) frequently numerical
simulations are the technique of choice (see for example \cite{BOOK}
and references therein).

Recently ground state techniques have become popular (see for example
\cite{RIEGER-GS} and references therein). The big advantage here is
that one is (down in the broken phase) surely as far as possible from
the critical temperature $T_c$. Our goal is to observe the effects of
the $T=0$ fixed point, and to remove spurious effects due to the
critical point at $T=T_c$: these contaminations are minimized when
working at $T=0$.  Also the use of exact ground states solves the
problem of thermalization, that is typically very severe on spin
glasses at low values of $T$. The price one pays is that contacts with
$T\ne 0$ physics are not clear: it could even be (but it does not seem
to be so, and here we will help in showing that this is not the case)
that what happens at $T=0$ is essentially different from what happens
for an even infinitesimal value of the temperature.

Here we will use a further tool to check consistency of the $T=0$ and
the $T\ne 0$ phase space. Thanks to off-equilibrium dynamical
simulations we will be able to study large lattice sizes and
restricted sectors of the phase space: this will allow to show better
that $T=0$ physics is compatible with the results one obtains at
finite values of the temperature.

The second important part of this work is that, always thanks to
off-equilibrium dynamical numerical simulations, we are able to keep
under control crucial observables like overlap-overlap correlation
functions and block overlaps. These observables are very important
since they allow to discriminate potentially misleading situations
like the ones where interfaces induce a seemingly non-trivial overlap
probability distribution $P(q)$ from a true replica symmetry
breaking. 

%%%%%%%%%%%%%%%%%%%%%%%%%%%%%%%%%%%%%%%%%%%%%%%%%%%%%%%%%%%%%%%%%%%%%%%%%%
\section{Model, Algorithm and Observables\label{S-MODEL}}

We have simulated a three dimensional ($3D$) Ising Spin Glass with
Gaussian couplings on a cubic lattice with periodic boundary
conditions. The Hamiltonian is
\begin{equation}
{\cal H} \equiv -\sum_{<i,j>} \sigma_i\  J_{ij}\  \sigma_{j} \ ,
\end{equation}
where the sum runs over all couples of nearest neighbors, the
$\sigma_i=\pm 1$ are Ising spins and the
couplings $J_{ij}$ are quenched random Gaussian variables with zero
mean and unit variance.

We have mainly focused on the measurements of two very important
observables: they are both useful to distinguish situations where a
true non-trivial behavior of the probability distribution of the
overlaps is present from situations where interfaces could present a
misleading situation (because of finite size effects or even in the
infinite volume limit, see the discussion in \cite{BLOCKS, MAPARIRUZU}
and references therein)

The first observable is an overlap, i.e. the measurements of how
similar two typical configurations at equilibrium are. Here we define
the overlap only in a small part of the lattice: we compute the block
overlap $q_{\cal B}$ on a $2 \times 2 \times 2$ cube (that we call
a $\cal B$-cube, here with ${\cal B} = 2$):
\begin{equation}
q_{\cal B}=\frac{1}{{\cal B }^3} \sum_{{\mathitbf i} \in {\cal B}}
q_{\mathitbf i}\, ,
\end{equation}
with the usual {site overlap}
\begin{equation}
q_{\mathitbf i} \equiv \sigma_{\mathitbf i} \tau_{\mathitbf i} ,
\end{equation}
where $\sigma $ and $\tau$ are two independent equilibrium
configurations of the system in the same realization of the quenched
disorder.  We call, as usual, $q$ the total overlap computed on all
the lattice, i.e. the average of the site overlap $q_i$ over all
lattice sites. We notice also that in the following we will, as usual,
denote with brackets the thermal average for one given realization of
the quenched disorder, $\langle \left(\ \cdot \ \right) \rangle$: when
averaging at a given Monte Carlo time $t$ we will mean averaging over
different realizations of the dynamical process, and we will indicate
this average with $\langle \left(\ \cdot \ \right) \rangle_t$.  We
will denote with an over-line the average over the quenched disorder,
$\overline{\left(\ \cdot \ \right)}$.

We have computed the full probability distribution $P_{\cal B}(q_{\cal
B},t,T)$ of the ${\cal B}=2$ block overlap $q_{\cal B}$ as a function
of the temperature $T$ and the (Monte Carlo) time $t$ of a so-called
{\em off-equilibrium dynamics}.  In an off-equilibrium numerical
simulation (see for example \cite{DYNAMIC_FIELD}) we use a ``very
large'' lattice size, and we work at values of the temperature low
enough to make sure that the system is not thermalized: this is true
already for medium size lattices in the case of system that are
characterized by slow dynamics, as is the case here. In this situation
we work by extrapolating off-equilibrium, finite time measurements to
infinite time. The advantage of the method is on one side in the fact
that we are basically free from finite size effects, and on the other
side that we can learn additional information from the pattern of the
approach to equilibrium (for example in \cite{DYNAMIC_FIELD} this turns
out to be a nice way to learn about the minimal value of the allowed
overlap, $q_{\mbox{min}}$). On the other side the disadvantage is that
we have to rely on an extrapolation to infinite time, that can turn
out to be not so trivial.

The second relevant observable is the overlap-overlap correlation
function (defined on the full lattice) computed at distance
${\mathitbf r}$, and it is defined as
\begin{equation}
  C({\mathitbf r},t,T) \equiv 
  \overline{ \langle q_{\mathitbf i} q_{{\mathitbf i}+{\mathitbf r}}   
  \rangle_t } \ .
  \label{E-C}
\end{equation}
Here we will only analyze $C(1,t,T)$, since among all the different
$C$ for different $\mathitbf r$ values it is the less affected by
statistical errors.

In our working conditions $C({\mathitbf r},t,T)$ depends on the
initial total overlap probability distribution $P(q,t=0,T)$.  In our
case, since we have used very large lattice sizes, since we are
starting from an average zero overlap (we start by evolving two
independent, uncorrelated random configurations) and since we are
limited to a number of Monte Carlo sweeps (of the full lattice) of the
order of $10^7$, we are not able to change away from zero the total
overlap, and at least roughly we have that $P(q,t,T) \simeq
P(q,t=0,T)$. Because of that we have that the correlation functions we
measure are very different from the equilibrium correlation functions,
and can be written as 
\begin{eqnarray}
  C({\mathitbf r},t,T)
  &=& \int dq\  P(q,t,T)\  C({\mathitbf r},t,T,q)\\
  \nonumber
  &\simeq& \int dq\  P(q,t=0,T)\  C_0({\mathitbf r},t,T,q)
\end{eqnarray}
where by $C({\mathitbf r},t,T,q)$ we denote the spatial correlation
functions at time $t$ and temperature $T$ computed by including only
couples of configurations having total overlap $q$, and by
$C_0({\mathitbf r},t,T,q)$ we denote the correlation functions
computed with initial conditions $P(q,t=0,T)$.

In the numerical simulations that we will discuss in the following we
have been starting from an initial probability distribution peaked
around zero total overlap, i.e. our results are determined by the
value of $C({\mathitbf r},t,T,q=0)$. In other terms we are discussing
here the dynamics of the $q=0$ sector of the system (see
\cite{MAPARIRUZU} for further details).  As we have stressed in the
introduction, our main goal here is to relate numerical results
computed from an off-equilibrium dynamics at non zero temperature with
equilibrium results obtained at zero temperature using techniques that
give exact ground state configurations on reasonably lattice sizes
(see for example~\cite{PAYO,MAPA}).

In the recent studies computing spin glass ground states the technique
of choice was based on computing the so-called {\em mixed overlap},
i.e. the overlap computed between a ground state in a finite volume
with periodic boundary conditions (pbc) in all directions and another
ground state obtained when imposing anti-periodic boundary conditions
(abc) in one direction and pbc in the other directions of the cubic
lattice with the same realization of the quenched couplings $J$.  In a
theory with continuously broken replica symmetry \cite{PARISI} in the
infinite volume limit there are many ground states with total site
overlap $q$ and total {\em link overlap} $q_l$ (that coincides with
the correlation function $C$ computed at distance 1) different from
one \cite{MAPA}. On the contrary a droplet approach \cite{DROPLET}
implies a behavior similar to the one of an usual ferromagnet, where
the link overlap (and the site overlap if we activate an infinitesimal
magnetic field) go to one in the thermodynamical limit. Both
\cite{PAYO} and \cite{MAPA} compute the ${\cal B }=2$ block overlap.

Now a key remark is in order. The procedure we have just described to
compute overlaps in the ground state sector leads to the comparison,
if the RSB picture holds, of ground states with $q=0$ (changing the
boundaries we favor ground states with small overlap): that means that
these results based on ground states describe the link overlap in the
$q=0$ sector of the system. This is the main reason of interest of
these dynamical simulations: we are simulating here
the $q=0$ sector of the theory, and since we are working on very large
lattice sizes we are only sampling this sector. In other words when
extrapolating our measurements of $C_\infty(1,T)$ (by the subscript
$\infty$ we mean we have already extrapolated, hopefully faithfully,
to $t=\infty$) from our low values of the temperature $T$
down to $T=0$ we should find, if the RSB picture is consistent, the same
results found in the direct analysis of the ground states.

We will also try to check the behavior of a different relevant
observable. In \cite{MAPA} the authors have computed the probability
that, for example, a ${\cal B} = 2$ block does not intersect an
interface: one can relate this probability to the probability of
finding that a ${\cal B}=2$ block is $\pm 1$.  Again our strategy will
be to compute the probability of this event at non zero temperatures
and to extrapolate the result down to $T=0$.

A very direct relation about the second moment of the (${\cal B}=2$)
block overlap and the $q-q$ correlation function \ref{E-C} can be
established by noticing that
\begin{equation}
\langle q_{\cal B}^2\rangle (t,T)=\frac{1}{8} \left[C(0,t,T)+ 3\ C(1,t,T) 
+ 3\ C(\sqrt{2},t,T) + C(\sqrt{3},t,T)\right] \ .
\label{eq:test}
\end{equation}

%%%%%%%%%%%%%%%%%%%%%%%%%%%%%%%%%%%%%%%%%%%%%%%%%%%%%%%%%%%%%%%%%%%%
\section{Numerical Results\label{S-NUMERICAL}}

Our numerical simulations have been run on a {\bf APE100} parallel
supercomputer \cite{APE}: on the more powerful version of the computer
(the so-called ``tower'' version) our Monte Carlo code for simulating
$3D$ Ising Spin Glasses with Gaussian couplings has a peak performance
of 5 GF.

For all the numerical simulations described in this note we have
averaged our data on four different realizations of the random
quenched disorder. In each disorder realization we have simulated two
independent copies of the system (useful to compute the different
overlaps, block overlaps and correlation functions). We have used a
straightforward Metropolis algorithm.  We have used a lattice of
linear size $L=64$ and volume $V=L^3=2^{18}$.

We have analyzed two kind of runs. ``Short'' runs have been used to
compute the extrapolated value of the overlap-overlap correlation
function. In these short runs we have simulated temperatures $T=0.9$,
$0.8$, $0.7$, $0.6$, $0.5$, $0.4$, $0.35$ and $0.3$ during
$819\;200$ Monte Carlo steps. We have presented a preliminary and
partial analysis of some of these data in reference \cite{COR} (only
for $T\ge 0.35$ and only as far the dependence of the critical
exponents over temperature is concerned: we had not analyzed, for
example, the prefactors, which are very relevant to the physics
studied in this paper, see below).

In the ``short'' runs we use a simple scheme. We start from two
initial independent random configurations, and suddenly start
iterating at the working temperature well below the critical one:
$T<T_c$ ($T=0.9$, $0.8$, $0.7$, $0.6$, $0.5$, $0.4$, $0.35$ or
$0.3$). At each $T$ value we will eventually extrapolate to infinite
time, to get a $T$ dependent asymptotic value, that we will, in turn,
try and extrapolate down to $T=0$.
 
In order to compute the dynamical behavior of the block overlap we
have used ``long'' runs. Here we have used the previous scheme for 
four values of the temperature, $T=0.7$, $0.6$, $0.5$ and
$0.35$, with $6\,553\,600$ Monte Carlo updates of the full lattice
performed at each value of $T$.

In the ``short'' runs we measure the time dependent observables on
logarithmic time scales: i.e. at times $t=100$, $200$, $400$,
$800$, \ldots In the ``long'' runs we measure every 32768 steps.

In the next two subsections we will discuss separately our numerical
results for the correlation functions and for the window overlap.

%%%%%%%%%%%%%%%%%%%%%%%%%%%%%%%%%%%%%%%%%%%%%%%%%%%%%%%%%%%%%%%%
\subsection{Correlation Functions\label{SS-CORRELATION}}

In this section we will always indicate by  $C(1,t,T)$ the $q=0$
component of the overlap-overlap correlation function, i.e. 
$C({\mathrm r}=1,t,T,q=0)$, that, as we have discussed in detail in
section \ref{S-MODEL}, is the quantity we are measuring.

The dependence of $C(1,t,T)$ on time can be expressed through the
knowledge of the dynamical critical exponent $z(T)$ (that depends on
$T$). We have that
$$
  C(1,t,T) = f\left(t^{\frac{1}{z(T)}},T\right)\ .
$$
From numerical simulations~\cite{RIEGER,Kisker,COR,COR1,COR2} 
(that here are in very nice agreement with real 
experiments~\cite{ORBACH}) we know that 
\begin{equation}
  z(T)=\frac1{T}\left(6.2\pm 0.3\right) \,  .
\label{EQ:ZETA}
\end{equation}
Another useful piece of information~\cite{COR,COR1,COR2} is that 
\begin{equation}
C(x,t,T)\propto  \frac{1}{\sqrt x} 
\exp\left[-\left(\frac{x}{\xi(t,T)}\right)^{3/2}\right]\, ,
\label{eq:Ansatz}
\end{equation}
where $\xi(T)\propto t^{1/z(T)}$ is the time
dependent {\em dynamical correlation length} of the system.
Because of that we expect that
\begin{equation}
\log C(1,t,T) \simeq a(T)+ b(T) t^{-3/\left(2 z(T)\right)}\ ,
\label{EQ:ANSATZ}
\end{equation}
asymptotically for large time.  As $t$ diverges $C(1,t,T)$ tends to
the constant value $\exp\left(a(T)\right)$, that we have called
prefactor in the previous section.

\begin{figure}[t!]
\begin{center}
\leavevmode
\epsfig{file=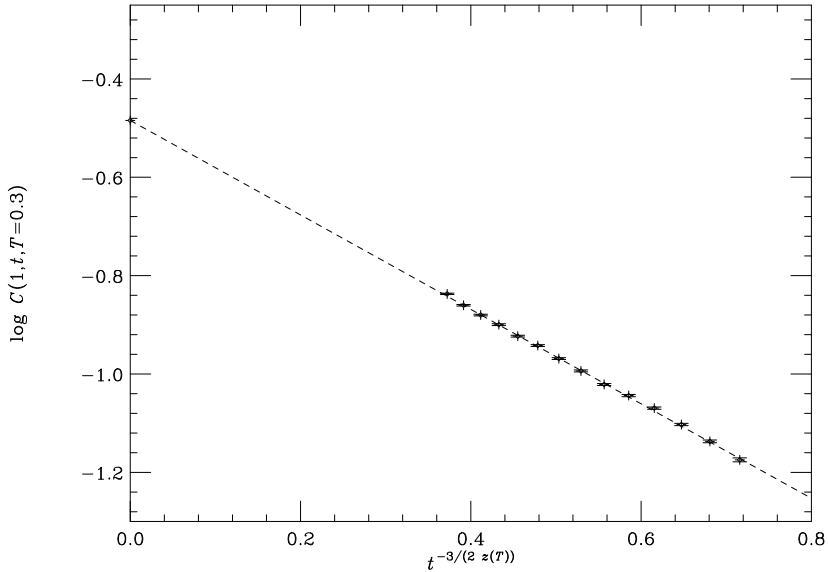,width=0.5\linewidth,angle=90}
\end{center}
\caption{
Values of $C(1,t,T)$ at $T=0.3$ (the lowest value of $T$ that
we use) versus $t^{-3/(2z)}$.
We also show our best fit to the form \ref{EQ:FIT}.}
\label{FIG:COR_LOW}
\end{figure}

\begin{figure}[t!]
\begin{center}
\leavevmode
\epsfig{file=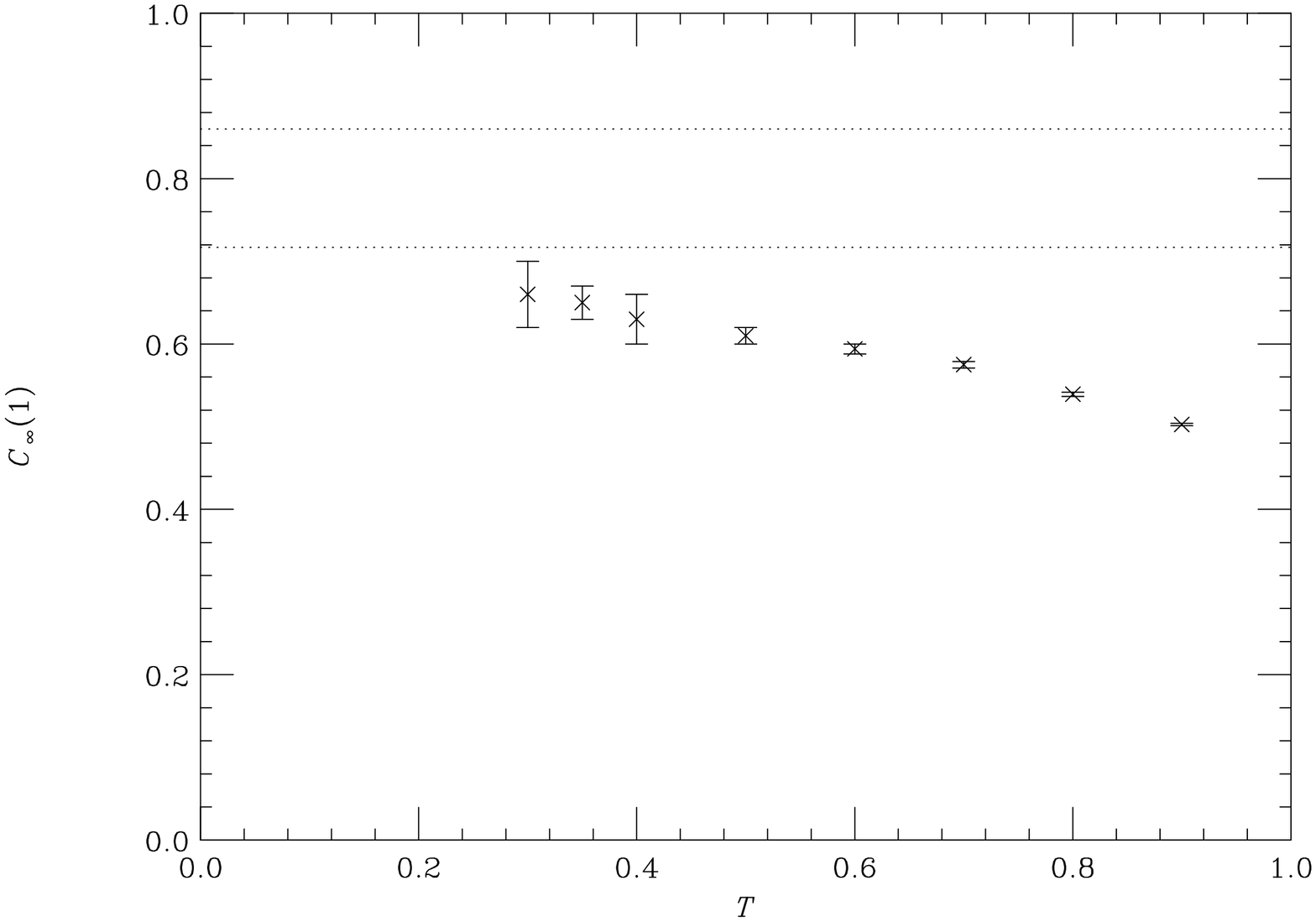,width=0.5\linewidth,angle=90}
\end{center}
\caption{Values extrapolated to infinite time of the correlation
function at distance $d=1$, $C_{\infty}(1,T)$, versus $T$. We have
also marked by two horizontal dotted lines the interval where the
value computed in reference~\cite{MAPA} using $T=0$ ground state
calculations lies.  The consistency of the two results is clear.}
\label{FIG:EXTRA}
\end{figure}

In figure \ref{FIG:COR_LOW} we show the extrapolation of $C(1,t)$ to
$t=\infty$ for the lowest temperature we use, $T=0.3$ (this is the
most difficult extrapolation, and fits at larger $T$ values are
easier). The dashed line is for the best fit to the form
\begin{equation}
\log C(1,t,T)= a(T)+ b(T)\  t^{-c(T)}\; .
\label{EQ:FIT}
\end{equation}
The fit is good, and the extrapolation looks reasonably safe. By this
kind of fits we are able to obtain reliable extrapolated values for
all the values of $T$ we study, that we show in figure \ref{FIG:EXTRA}
versus the temperature $T$.  We have checked that the value of $c(T)$
of the best fit to the form \ref{EQ:FIT} is in very good agreement
(i.e. inside the margins suggested by the estimated statistical error)
with the value of $3/(2 z(T))$. The discrepancies between the
extrapolated value obtained by using a three or two parameter fit
(i.e. by fixing $c(T)$ to $3/(2 z(T))$ in equation \ref{EQ:FIT} or
by letting it free) are inside the statistical errors.

For a detailed numerical study of the functional form suggested in
equation \ref{EQ:FIT} we refer the reader to
references~\cite{COR,COR1,COR2}. An exhaustive analysis of the
correlation functions computed at zero total overlap has been
performed in three~\cite{COR,COR1,COR2}, four~\cite{4D} and six
dimensions~\cite{6D}: in all cases the system turns out to have a very
similar behavior, independently of the spatial dimensionality. In six
dimensions the agreement with the quantitative analytical predictions
obtained by De Dominicis, Kondor and Temesvari~\cite{DEDO} is very
good~\cite{6D}.

In the same figure we have also plotted the statistical interval
allowed for the value of the link overlap at $T=0$ from ground state
calculations (see~\cite{MAPA}).  The statistical error on our values
of $C_{\infty}(1,T)$ grows when $T$ decreases, but in the limit given
by these errors the $T=0$ result and a reasonable extrapolation of the
finite $T$ results obtained here look completely consistent.

The interval we have drawn in the figure is $0.79\pm 0.07$, and we
have computed it by looking at data from~\cite{MAPA}. We have used the
four different (but statistically compatible) values obtained there
for $q_l$. By extrapolating of $q_l$ for $L\to\infty$ one obtains
$q_l=0.755\pm 0.015$ or $q_l=0.80\pm 0.06$ using a linear or a
quadratic fit in $1/L$, respectively.  By studying correlation
function reference~\cite{MAPA} also obtained $C(1)=0.732\pm 0.008$
(for the transverse correlation) and $C(1)=0.722\pm 0.005$ (for the
perpendicular correlation). Taking into account all these figures we
have obtained the confidence interval (at one standard deviation) for
the link overlap that we were quoting before.

%%%%%%%%%%%%%%%%%%%%%%%%%%%%%%%%%%%%%%%%%%%%%%%%%%%%%%%%%%%%%%%%%%%%%%%%%%
\subsection{Block Overlaps\label{SS-BLOCK}}

Our second set of measurements concerns the so called {\em block
overlap} (that we also call {\em window overlap} or {\em box overlap},
and have been computed on a ${\cal B} \times {\cal B} \times {\cal B}$
$=$ $2 \times 2 \times 2$ box). As in the case of correlation
functions we start from two independent random configurations, and
cool them down according to the annealing schedule we have described
before (using what we have called the ``long'' runs). During these
runs we have computed the full probability distribution of the window
overlap at different times $t$.

In this section of this note our main goal is to compute $P_{\cal
B}(\pm 1)$, i.e.  the value (extrapolated to infinite time first and
than $T=0$ later) of the probability that a $\cal B$ block has overlap
of maximal modulus. To simplify the notation from now on we will be
working with a symmetrized overlap probability distribution, i.e. we
will consider $P_{\cal B}(|q|,t,T)$, that can have support in $(0,1)$:
this object is interesting because it is exactly the probability that
the interface between different phases does not intersect with a cube
of size $2 \time 2 \time 2$. As we have already discussed this
probability has been computed in reference~\cite{MAPA} at $T=0$, and
we will try here to connect these results with the $T\ne 0$ physics.

The main result of reference~\cite{MAPA} is that, at $T=0$ and in the
limit of the lattice size $L\longrightarrow \infty$,
\begin{equation}
  P_1({\cal R}) \simeq  0.65 \pm 0.05\ .
\end{equation}
Our working Ansatz for the time dependence of $P_{\cal B}(|q|=1)$ has been,
as we will justify now, that
\begin{equation}
\log P_{\cal B}(|q|=1,t,T)=\log P_{\cal B}^\mathrm{t=\infty}(|q|=1,T)
+\frac{a(T)}{t^{3/(2 z(T))}} \  ,
\label{pqscaling}
\end{equation}
and  we have used $z(T)=\frac{6.2}{T}$.

Let us see why things work in this way. From equation \ref{eq:test}
we know that $<q^2>_B$ scales as an overlap-overlap correlation
function. But for large times, and for distances such that $x \ll
t^{1/z}$ 
equation
\ref{eq:Ansatz} tells us that
the correlation function behaves as
\begin{equation}
C(x,t) \propto \frac{1}{\sqrt x} \left[1- A \frac{x}{t^{3/(2 z(T))}}
\right] \; ,
\end{equation}
where $A$ is a suitable constant. This implies in turn that
\begin{equation}
<q_{\cal B}^2>(t,T) \simeq a(T) +\frac{b(T)}{t^{3/(2 z(T))}} \ ,
\label{eq:extra}
\end{equation}
where $a(T)$ and $b(T)$ are constants that only depend on $T$.
Now since 
\begin{equation}
<q_{\cal B}^2>(t,T) \equiv \int_{-1}^1 dq \, P_{\cal B}(q,t,T)~ q^2 \; ,
\end{equation}
if we define $\Delta P_{\cal B}(q,t,T) \equiv P_{\cal B}(q,t,T)-P_{\cal
B}^\mathrm{t=\infty}(q,T)$ it is clear that
\begin{equation}
\int_{-1}^1 dq \, \Delta P_{\cal B}(q,t,T)~ q^2\simeq \frac{b(T)}{t^{3/(2
z(T))}} \; .
\label{eq:q2t}
\end{equation}
Things are more clear if we write equation \ref{eq:q2t} for discrete
values of $q_{\cal B}$ (that is what happens for finite $\cal B$):
\begin{equation}
\sum_{ q < 1}  \Delta P_{\cal B}(q,t,T)~ q^2 
+ \Delta P_{\cal B}(q=1,t,T) = \frac{b(T)}{t^{3/(2
z(T))}} \; .
\label{eq:q2t_dis}
\end{equation}
If we assume that all the components with $q>0$ scale 
in the same way this shows that
\begin{equation}
  \Delta P_{\cal B}(q=1,t,T) \equiv
  P_{\cal B}(q=1,t,T)-P_{\cal B}^\mathrm{t=\infty}(q=1,T)
  \simeq t^{-3/(2 z(T))} \;.
\label{eq:scaling_final}
\end{equation}
and so, 
\begin{equation}
P_{\cal B}(|q|=1,t,T)= P_{\cal B}^\mathrm{t=\infty}(|q|=1,T)
+\frac{d(T)}{t^{3/(2 z(T))}} \  ,
\label{eq:pqscaling_b}
\end{equation}
and we recover \ref{pqscaling} taking logs in \ref{eq:pqscaling_b}. We
have used \ref{pqscaling} instead of \ref{eq:pqscaling_b} since we
have (slightly) better scaling even for small times. We have checked that both
kind of fits provide us with compatible values of $P_{\cal
B}^\mathrm{t=\infty}(|q|=1,T)$.

We will use equation \ref{pqscaling} to fit the values of
$P_{\cal B}^\mathrm{t=\infty}(q=1,T)$ at different temperatures. The
argument we have done says nothing about the behavior of $\Delta
P_{\cal B}(q=0,t,T)$, while $\Delta P_{\cal B}(q,t,T)$ for $q \neq 0$
follows the same scaling law valid for $\Delta P_{\cal B}(q=1,t,T)$.

\begin{figure}[t!]
\begin{center}
\leavevmode
\epsfig{file=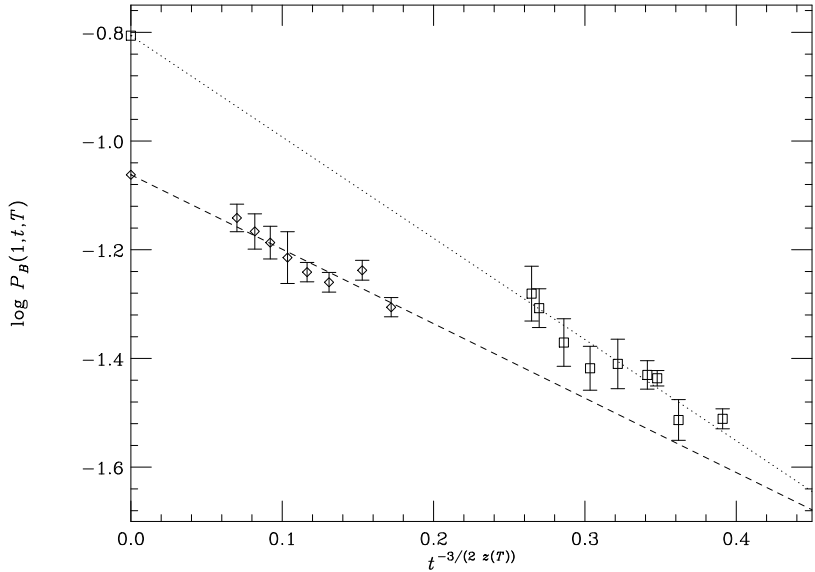,width=0.5\linewidth,angle=90}
\end{center}
\caption{
$\log P_{\cal B}(1,t,T)$ versus $t^{-\frac{3}{2 z(T)}}$ for
$T=0.7$ (rhombs) and $T=0.35$ (squares).
The dashed line (dotted line) is for the best fit to the Ansatz
\ref{pqscaling} for $T=0.7$ ($T=0.35$ respectively).}
\label{FIG:pq1_extra}
\end{figure}

We have checked that our numerical data are well fitted by the
behavior of equation \ref{pqscaling}. We show two plots corresponding
to two different temperatures: $T=0.7$ and $T=0.35$, in figure
\ref{FIG:pq1_extra}. In the plots we also draw our best fits to to the
Ansatz \ref{pqscaling}.

\begin{figure}[t!]
\begin{center}
\leavevmode
\epsfig{file=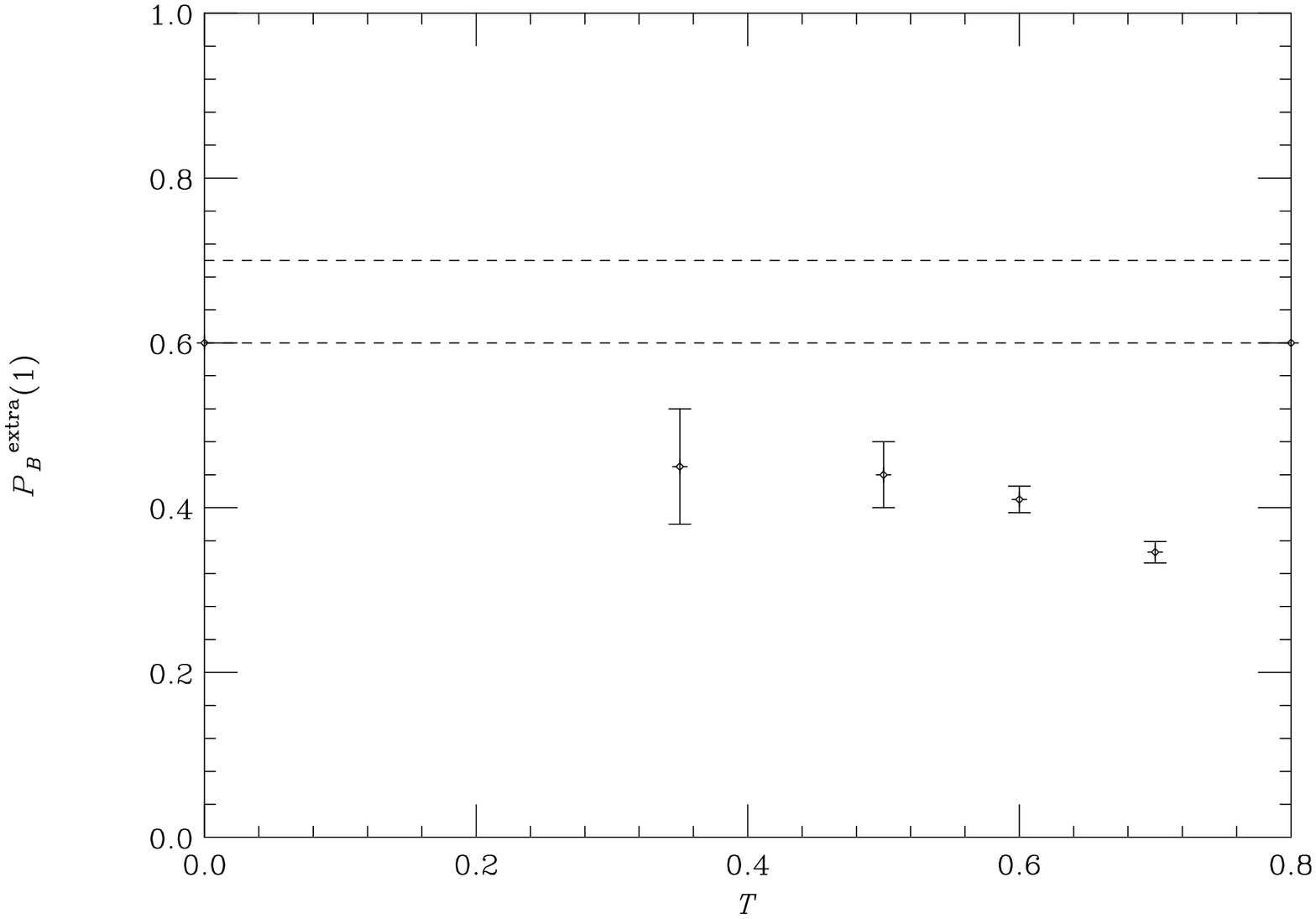,width=0.5\linewidth,angle=90}
\end{center}
\caption{$P_{\cal B}^\mathrm{t=\infty}(1,T)$ versus $T$.  The
horizontal line corresponds to the confidence limit obtained in ground
state computations.  }
\label{FIG:extra_pq_1}
\end{figure}

The fit of figure \ref{FIG:pq1_extra} for $T=0.7$, for example, looks
reasonably reliable. The $P_{\cal B}$ depends very slowly on $T$, and
our numerical data have as support a range of times as large as the
range we need to extrapolate through.  On the contrary the fit of
\ref{FIG:pq1_extra} for $T=0.35$ (again, the one at our lowest $T$
value, i.e. our most difficult fit) looks less happy even if one would
not expect very strange things to happen: since $\frac1{z(T)}$ at such
low $T$ is becoming very small our data vary on a small support, and
the $t=\infty$ point looks far away. The very large error that appears
in figure \ref{FIG:extra_pq_1} is an effect of this phenomenon.

In figure \ref{FIG:extra_pq_1} we show the values of $P_{\cal
B}^\mathrm{t=\infty}(q=1,T)$ extrapolated up to $t=\infty$ versus
$T$. 
Because of how difficult it is to get precise results at low $T$ values,
the evidence we have established is not very strong: still, figure
\ref{FIG:extra_pq_1} clearly shows that our present low $T$ results
are completely compatible with the $T=0$ results.

We end this section by performing a last consistency check. We compare
the values of $\langle q_{\cal B}^2\rangle$ with the values of the
correlation function (in both cases extrapolated to infinite time).
Using equation \ref{eq:Ansatz} we can write equation \ref{eq:test} as:
\begin{equation}
  \langle q_{\cal B}^2\rangle(T) = 0.125 + 0.785 ~ C_{\infty}(1,T) \;.
  \label{finaltest}
\end{equation}
In table \ref{table1} we show the values of $\langle q_{\cal
B}^2\rangle(T)$ obtained by direct measurements of $q_{\cal B}$ and
the value obtained by summing up correlation functions, according to
the right hand side of equation \ref{finaltest}: the agreement of the
two quantities is very reasonable at all $T$ values.  The values of
$\langle q_{\cal B}^2\rangle(T)$ have been extrapolated at infinite
time by using equation \ref{eq:extra}.

\begin{table}
\begin{center}
\begin{tabular}{|c|c|c|}\hline
T    & $\langle q_{\cal B}^2\rangle(T)$ & rhs of
equation  \ref{finaltest}\\ \hline
$0.7  $ & $0.585\pm 0.002		$ & $0.576\pm 0.004 $ \\ \hline
$0.6  $ & $0.596\pm 0.009		$ & $0.594\pm 0.004 $ \\ \hline
$0.5  $ & $0.57\pm 0.02             $ & $0.60\pm 0.01   $ \\ \hline
$0.35 $ & $0.61\pm 0.03		$ & $0.64\pm 0.02   $ \\ \hline
\end{tabular}
\caption{Test of relation \ref{finaltest}.}
\label{table1}
\end{center}
\end{table}

%%%%%%%%%%%%%%%%%%%%%%%%%%%%%%%%%%%%%%%%%%%%%%%%%%%%%%%%%%%%%%%%%%%%%%%%%%
\section{Conclusions\label{S-CONCLUSION}}

The results of our investigation are indeed positive. Physical
observables computed at $T=0$ are compatible with values computed on
large lattices thanks to an off-equilibrium dynamics, and extrapolated
from measurements at low, finite $T$. Our observables of choice are
interesting observables, since they do not only show that our system
has a non-trivial $P(q)$, but also that this non-trivial $P(q)$ is the
``bona fide'' effect of replica symmetry breaking, and not the effect
of some fancy kind of interfaces.

We have been able to use very low $T$ values, while still keeping a
good control over $t\to\infty$ fits.  When we lower $T$ and eventually
$T\longrightarrow 0$ the data are collapsing, in the reasonable
accuracy given by the large statistical error, to the $T=0$ results.
It is important to remark, also as far as planning future, more
precise studies is concerned, that it is really very difficult to run
accurate numerical simulations in the region $T\le 0.3$ with the
nowadays computers (for example at $T=0.1$ the dynamical critical
exponents is of the order of $60$).

%%%%%%%%%%%%%%%%%%%%%%%%%%%%%%%%%%%%%%%%%%%%%%%%%%%%%%%%%%%%%%%%%%%%%%%%%%
\section*{Acknowledgments}

JJRL acknowledges partial financial support from CICyT (PB98-0842).

%%%%%%%%%%%%%%%%%%%%%%%%%%%%%%%%%%%%%%%%%%%%%%%%%%%%%%%%%%%%%%%%%%%%%%%%%%

\end{document}